\documentstyle[epsf,aps,twocolumn,epsfig]{revtex}

\begin{document}
\title{Properties of cage rearrangements observed near the colloidal
glass transition}
\author{Eric R.~Weeks$^{*, \rm (1)}$ and D.~A.~Weitz$^{\rm (2)}$}
\address{$^{\rm (1)}$Physics Department, Emory University, Atlanta, GA 30322}
\address{$^{\rm (2)}$Department of Physics and DEAS, Harvard University,
Cambridge, MA 02138}
\date{February 4, 2002}
\maketitle

\begin{abstract}
We use confocal microscopy to study the motions of particles
in concentrated colloidal systems.  Near the glass transition,
diffusive motion is inhibited, as particles spend time trapped
in transient ``cages'' formed by neighboring particles.
We measure the cage sizes and lifetimes, which respectively
shrink and grow as the glass transition approaches.  Cage
rearrangements are more prevalent in regions with lower local
concentrations and higher disorder.  Neighboring rearranging
particles typically move in parallel directions, although a
nontrivial fraction move in anti-parallel directions, usually
from pairs of particles with initial separations corresponding
to the local maxima and minima of the pair correlation function
$g(r)$, respectively.
\end{abstract}

%\pacs{PACS: 61.43.Fs (glasses), 64.70.Pf (glass transitions),
%82.70.Dd (colloids), do we want one on microscopy?
%61.20.Ne (structure of simple liquids)}

Many liquids undergo a glass transition when rapidly cooled, where
their viscosity grows by orders of magnitude for only modest
decreases in temperature.  This drastic increase in viscosity is
unaccompanied by significant structural changes; instead, the
dynamics slow dramatically.  Physically, this slowing of the
dynamics reflects the confinement of any given particle by a
``cage'' formed by its neighbors; it is the rearrangement of the
cage which leads to the final structural relaxation, allowing the
particle to diffuse through the sample \cite{reviews}.  The
dynamics of cages have been studied with scattering measurements,
which probe a spatial and temporal average of their behavior, and
with computer simulations; however, in real systems, the actual
motion of the individual particles involved in cage dynamics and
breakup are still poorly
understood~\cite{reviews,kegel,weeks,kob97,donati98,donati99,perera99,berne,doliwa98}.

\begin{figure}
\centerline{
\epsfxsize=8.0truecm
\epsffile{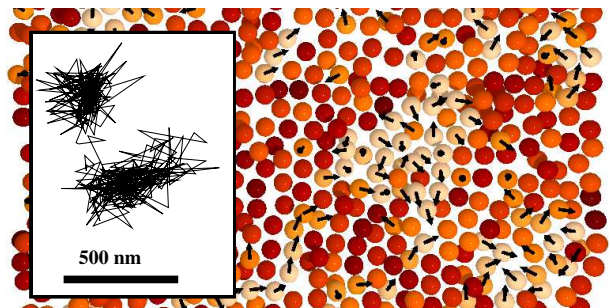}}
\smallskip
\caption{Cut through a three-dimensional sample,
with arrows indicating the direction of motion for particles with
displacements $\Delta r > 0.2$ $\mu$m, using $\Delta t^* = 600$~s.
The sample has $\phi=0.52$, and the cut is 2.5 $\mu$m thick ($\sim
1$ layer of particles).  The arrows are all the same length in
three dimensions, so
shortened arrows indicate motion in or out of the picture.
Lighter colors indicate particles with larger displacements.
Inset:
120 min trajectory of one particle from this sample. }
\label{picture}
\end{figure}

In this paper, we study the motion of the individual particles
and their neighbors during cage breakup, and provide the first
direct experimental visualization of this process.
We use confocal microscopy to study the motion of colloidal
particles in a dense suspension, an excellent model for the
glass transition~\cite{pusey86,bartsch93,vanmegen94}.
The rearrangement of cages involves the
cooperative motion of neighboring particles
\cite{kegel,weeks,kob97,donati98,donati99,perera99}, for example as shown in
Fig.~\ref{picture}, where the most mobile particles
have arrows indicating the direction of their motion.
While most neighboring particles move in similar directions,
a significant minority move in opposite directions, resulting
in local changes in topology.  We find particles moving in parallel
directions typically have initial separations corresponding
to local maxima of the pair correlation function $g(r)$,
while pairs of particles moving in anti-parallel directions
typically correspond to local minima of $g(r$).  We also find
that the more mobile particles are located in regions with
a lower local volume fraction, and higher disorder.  These
measurements provide a direct, quantitative physical picture
of the nature of cage rearrangements.

% PICTURE - FIG
\smallskip

We use sterically stabilized poly-(methylmethacrylate) particles with a
radius $a=1.18$ $\mu$m \cite{weeks,pusey86,dinsmore}, immersed in a
mixture of decalin and cycloheptylbromide.  This solvent
simultaneously matches the particle index of refraction and
density to
mitigate the effects of scattering and sedimentation.  Hard
sphere particles undergo a freezing transition at a volume fraction
$\phi_{f}=0.494$, a melting transition at $\phi_m=0.545$, and have a
glass transition at $\phi_g \approx 0.58$ \cite{pusey86,vanmegen94}. To
visualize the particles, we stain them with fluorescent rhodamine dye;
this imparts a slight charge to the particles, shifting the phase
transition boundaries to $\phi_f \approx 0.38$, $\phi_m \approx 0.42$,
and $\phi_g \approx 0.58$. We image a volume 60 $\mu$m $\times$ 60
$\mu$m $\times$ 10 $\mu$m, containing several thousand particles, and
identify particle centers with an accuracy of at least 0.05 $\mu$m
\cite{dinsmore,crocker}. A typical particle trajectory is shown in the
inset of Fig.~\ref{picture}; it exhibits caged motion, with a sudden
cage rearrangement which lasts $\sim$600~s.

\begin{figure}
\centerline{ \epsfxsize=8.0truecm \epsffile{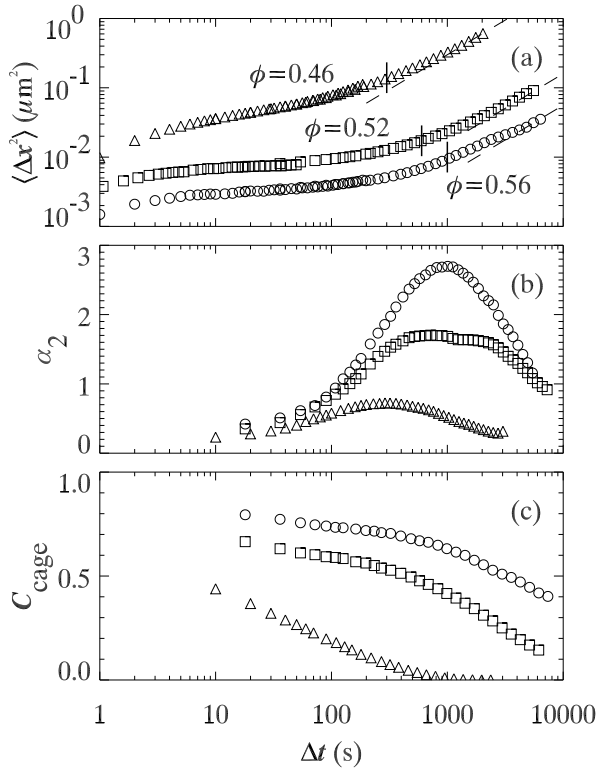}}
%\centerline{ \epsfxsize=12.0truecm \epsffile{fig2.ps}}
\smallskip
\caption{(a) Mean square displacements.
%%%%>> $\langle \Delta x^2 \rangle$
The diagonal straight lines indicate estimates for $
\langle \Delta x^2 \rangle \sim 2 D_{\infty} \Delta t$.
Vertical bars indicate $\Delta t^*$.
$\langle \Delta x^2 \rangle$ is shown rather than
$\langle \Delta r^2 \rangle$ as the poorer $z$-resolution
of the microscope artificially increases $\langle \Delta r^2
\rangle$.
(b) Nongaussian parameter $\alpha_2(\Delta t)$.
(c) The cage correlation
function.} \label{time}
\end{figure}

The effect of cages on the ensemble dynamics is well known
\cite{reviews,kegel,weeks,kob97,donati98,donati99,perera99,berne,doliwa98,vanmegen94}
and is evident from the particle mean
square displacement, $\langle\Delta x^2 \rangle$, shown in
Fig.~\ref{time}(a) for three supercooled colloidal fluids ($\phi
< \phi_g$).  At the earliest time scales, particle motion
is diffusive, as they have not moved far enough to encounter
the cage formed by their neighbors.
As their displacement becomes larger, their motion is impeded
by the cage, leading to the plateau in $\langle\Delta x^2
\rangle$. The displacement at the plateau decreases as $\phi$
increases, reflecting the smaller cage size.  Moreover, the
cages become more long-lived with increasing $\phi$; this
presumably results from the fact that cage rearrangements
involve a larger number of particles as the glass transition
is approached~\cite{kegel,weeks,kob97,donati98,donati99,perera99}. The
cage rearrangement itself leads to an upturn in $\langle\Delta
x^2 \rangle$ at the end of the plateau on the log-log
plot; at even longer lag times, the asymptotic motion again
becomes diffusive, albeit with a greatly decreased diffusion
coefficient, $D_\infty$, as indicated by the dashed lines
in Fig.~\ref{time}(a).

% TIME - FIG
\smallskip

\begin{figure}
%\centerline{ \epsfxsize=10truecm \epsffile{fig3.ps}}
\centerline{ \epsfxsize=8.0truecm \epsffile{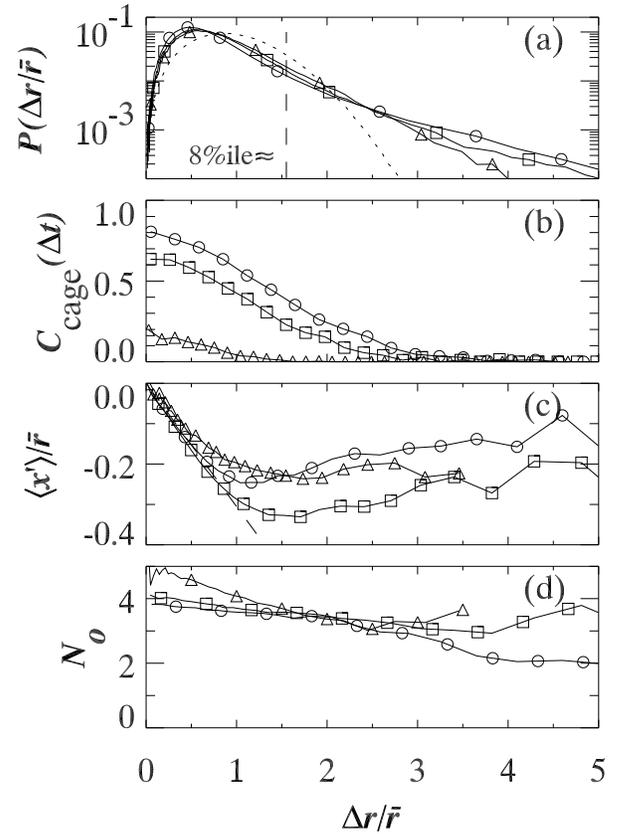}}
\smallskip
\caption{ (a) Step-size distribution. The vertical dashed line
indicates the 8\%ile cutoff.  The dotted line is a gaussian with
width $\bar r = 1$. (b) $C_{\rm cage}(\Delta t^*;\Delta r)$. (c)
Average subsequent displacement $\langle x' \rangle$ along the direction
of the original displacement $\Delta r$.  The dashed line is a
linear fit to the small $\Delta r$ region for the $\phi=0.56$ data
(circles).
(d) The average number of
ordered neighbors $N_o$ particles have. The symbols are: triangles
$\phi=0.46$, $\Delta t^*=300$ s, $\bar r = 0.61$ $\mu$m; squares
$\phi=0.52$, $\Delta t^*=600$ s, $\bar r = 0.22$ $\mu$m; circles
$\phi=0.56$, $\Delta t^*=1000$ s, $\bar r = 0.16$ $\mu$m.}
\label{speed}
\end{figure}

We estimate the time scale for cage rearrangement, $\Delta t^*$,
by finding the maximum of the nongaussian parameter
$\alpha_2(\Delta t) = \langle \Delta x^4 \rangle / (3 \langle
\Delta x^2 \rangle) - 1$, shown in Fig.~\ref{time}(b).  $\alpha_2$
is computed from the 1D displacement distribution function
$P(\Delta x;\Delta t)$ \cite{weeks,kob97,donati98,donati99}, and
is zero for a gaussian distribution, and largest when there are
broad tails.  To compare these distributions for different $\phi$,
we normalize the displacements $\Delta r$ by the average, $\bar r
\equiv \langle\Delta r^2 \rangle^{1/2}$.  The radial step size
distribution $P(\Delta r / \bar r; \Delta t^*)$ is plotted for
three values of $\phi$ in Fig.~\ref{speed}(a).  There are more
large displacements than expected for a gaussian distribution,
shown by the dashed line; moreover, as $\phi_g$ is approached, the
tails become even broader, as shown by the circles for $\phi =
0.56$.  This reflects the anomalously large motion of the
particles undergoing cage rearrangements.

% SPEED - FIG
\smallskip
%% (d) The average Voronoi volume per particle $V$,
%% normalized by the sample average $V_0$.

These large displacements contribute to the increase in the mean
square displacement at lag times greater than $\Delta t^*$.  To
confirm that this motion directly reflects structural relaxation,
we calculate a topological cage correlation function, $C_{\rm
cage}(\Delta t)$~\cite{berne}. We define particles as nearest
neighbors if their separation is less than a cutoff distance set
by the first minimum of $g(r)$; $C_{\rm cage}(\Delta t)$ is the
fraction of particles with the same neighbors at time $t$ and time
$t + \Delta t$, averaged over all $t$.  As shown in
Fig.~\ref{time}(c), for the lowest volume fraction, $\phi = 0.46$
(triangles), the particles are barely caged, but as $\phi$
increases toward $\phi_g$, the decay of $C_{\rm cage}$ slows
dramatically.  To determine which particles are responsible for
topological changes, we replot $C_{\rm cage}(\Delta t^*)$ as a
function of the normalized displacement.  As shown in
Fig.~\ref{speed}(b), $C_{\rm cage}(\Delta t^*)$ decreases
significantly for particles with larger displacements, confirming
that these particles contribute most to the structural relaxation.

To directly measure the size of the cage, we investigate the
temporal correlations of the motion of individual
particles~\cite{doliwa98}. Particles that remain caged must have
no significant net displacement over long times; by contrast those
whose cages rearrange do have net displacements.  To quantify
this, we compare a particle's displacements $\Delta \vec r$ and
$\Delta \vec r\,'$ during sequential time intervals of $\Delta
t^*$.  We focus on $x'$, the component of $\Delta \vec r\,'$ along
the direction of the original displacement $\Delta \vec r$;
$\langle x' \rangle$ is plotted in Fig.~\ref{speed}(c).  $\langle
x' \rangle$ is always less than zero, indicating that the average
motion is anticorrelated.  For small initial displacements the
behavior is linear, $x' = -c |\langle \Delta \vec r \rangle|$;
larger values of $c$ indicate more highly anticorrelated motion
\cite{doliwa98}. The cage constrains a particle so that the
farther a particle moves, the farther it will move back.  The
linear relationship fails for larger displacements, as seen in
Fig.~\ref{speed}(c); thus the degree of anticorrelation actually
decreases -- particles move shorter distances $\langle x' \rangle$
than expected by linear extrapolation from the small $\Delta \vec
r$ behavior. We identify the end of the linear regime as the cage
size, $r_{\rm cage}$ \cite{doliwa98}. For comparison, we estimate
the size of the cage by other methods.  The simplest is based on
the free volume of the system compared to random close packing,
$r_{\rm free} = 4a [ (\phi_{rcp}/\phi)^{1/3} -1 ]$. Another
estimate comes from the mean square displacement at the cage
rearrangement time scale $\Delta t^*$, $r_{\rm msd} = \langle
\Delta r^2(\Delta t^{*}) \rangle^{1 \over 2}$. The final estimate
is $\Delta r^{*}(\Delta t^*)$, chosen so that 5\% of particles
have displacements $\Delta r > \Delta r^*$ ~\cite{weeks,donati99}.
The values of these estimates are listed in Table 1 and are in
good agreement.  The cage size decreases as the glass transition
is approached, in agreement with expectations
\cite{doliwa98,allegrini99}.

We can use this cage size to model the particle motion as a random
walk, alternating between steps (cage rearrangements) and pauses
(caging). At long times, the diffusive motion reflects particles
undergoing many steps of size $r_{\rm cage}$ in random directions.
>From the Central Limit Theorem, $D_\infty = r_{\rm cage}^2 / 2
\Delta t^{**}$, where $\Delta t^{**}$ is the mean time between
steps, or the average cage lifetime \cite{weeks98}.  The measured
values of $D_\infty$, and the values calculated for $\Delta
t^{**}$ are listed in Table 1.  We find that $\Delta t^{**}$ is
significantly larger than $\Delta t^*$, the time scale for a
particle to move {\it during} one cage rearrangement.  Both grow
as $\phi$ increases, consistent with previous experiments
\cite{reviews,bartsch93,vanmegen94} and simulations
\cite{kob97,donati98,donati99,perera99,berne,doliwa98,allegrini99,donati99}.
While the cage size decreases, it remains finite as $\phi_g$ is
approached, implying that the dramatic decrease in $D_\infty$ is
due primarily to the increasing cage lifetime.  Our random-walk
picture also yields an unambiguous estimate of the fraction of
particles involved in cage rearrangements at any given time; it is
given by the ratio $\Delta t^* / \Delta t^{**}$.  This ratio is
$\sim$8\% for all samples except the most dilute (for which it is
16\%); it is shown by the vertical dashed line in
Fig.~\ref{speed}(a).

Several computer simulations looked for the underlying
origins of cage rearrangements, and found correlations
between mobile particles and their local environments
\cite{donati99,perera99,bulbul01,luchnikov95}.  Mobile
particles tended to be in regions with higher disorder
\cite{perera99,bulbul01,luchnikov95}, lower density
\cite{perera99,luchnikov95}, and higher potential energy
(and thus higher disorder) \cite{donati99}.  Ultimately,
our colloidal samples will crystallize, and it is possible
that the evolution from disorder to crystalline order drives
the structural relaxation in our experiments.  To investigate
this possibility, we quantify the local order with an order
parameter that identifies local crystalline regions from
particle positions \cite{wolde}.  If two adjacent particles
have similar orientations of their neighbors about each of
them, the two particles are ordered neighbors \cite{bonds};
most particles have between 2 and 4 ordered neighbors.
Particles with larger displacements have fewer ordered
neighbors, as shown in Fig.~\ref{speed}(d), in agreement
with the correlation between mobility and disorder seen in
simulations \cite{donati99,perera99,bulbul01,luchnikov95}.
Furthermore, the mobile colloidal particles typically move to
positions where their number of ordered neighbors has increased
by $\sim$1, suggesting a slow evolution toward crystalline
structure.  We emphasize, however, that crystallization occurs
on time scales significantly longer than these observations;
moreover, crystallization is a nucleation and growth process,
with full crystalline order of the sample resulting from the
growth of only a small number of nuclei.  Thus, any effect
of the local crystal order on the cage rearrangement may be
driven more by local variations in volume fraction, rather
than an evolution to the state with the lowest global energy
minimum \cite{wolde}.  In fact, we find that the volumes of
the Voronoi polyhedra associated with the mobile particles are
larger on average, giving a local volume fraction as much as
$\delta \phi=O(0.03)$ lower than the mean.  This suggests that
the particles with smaller $\phi$ are farther from $\phi_g$
and thus are more likely to rearrange, in agreement with the
simulations \cite{donati99,perera99,luchnikov95}.

%% commented out by Eric:
%% The variation in local volume fraction is $\delta \phi =
%% O(0.03)$, which can be quite significant, particularly close to
%% $\phi_g$, where small variations in $\phi$ can cause dramatic
%% changes in behavior.  After the particles have moved, their
%% Voronoi volume is generally closer to the average.

\begin{figure}
%\centerline{ \epsfxsize=10.0truecm \epsffile{fig4.ps}}
\centerline{ \epsfxsize=8.0truecm \epsffile{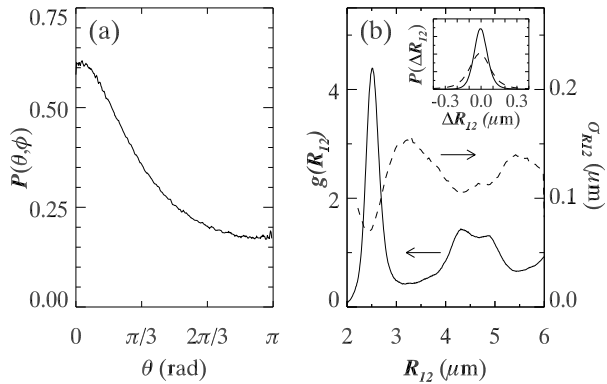}}
\smallskip
\caption{(a) Probability distribution of angle between displacement
vectors for two neighboring particles, as a function of the polar angle
$\theta$. (b) Pair correlation function $g(R_{12})$ (solid line) and
width $\sigma_{R12}$ of probability distribution function $P(\Delta
R_{12}|R_{12})$ (dashed line). Insert shows $P(\Delta R_{12}|R_{12})$
for nearest neighbors, with initial separations of 2.4 $\mu$m $<R_{12}<
2.6$ $\mu$m (solid line), and anti-neighbors with 3.0 $\mu$m $<R_{12}<
3.2$ $\mu$m (dotted line). All data are for $\phi=0.56$ and $\Delta t^*
= 1000 s$.} \label{separation}
\end{figure}

Cage rearrangement is not strictly a localized event; instead,
many neighboring particles are typically involved; they often move
in similar directions, as shown by the arrows in
Fig.~\ref{picture}
\cite{kegel,weeks,kob97,donati98,donati99,perera99}. To quantify
the propensity for motion of neighboring particles in the same
direction, we calculate the distribution of angles, $P(\theta)$,
between the displacement vectors of all neighboring particles,
measured at $\Delta t^*$ to define the displacements.  The
probability of observing two displacement vectors forming an angle
in the range $(\theta,\theta+d\theta;\phi,\phi+d\phi)$ is given by
$P(\theta,\phi)\sin\theta\, d\theta\, d\phi$; in
Fig.~\ref{separation}(a) we plot $P(\theta,\phi)$ which, by
symmetry, depends only on $\theta$. This function is strongly
peaked at $\theta \approx 0$, indicating that two particles
usually move in parallel directions (``strings'')
\cite{donati98,donati99,perera99}, although a significant fraction
of particles move in anti-parallel directions ($\theta \approx
\pi$, ``mixing regions'') \cite{footnote}. When $\theta \approx
\pi$, particles are either converging or diverging; examples of
each case can be seen in Fig.~\ref{picture}.  We find that these
``mixing'' pairs of particles ($\theta > \pi/2$) are approximately
twice as likely to result in the two particles no longer being
neighbors (compared to the pairs with with $\theta < \pi/2$).
Thus, these rearrangements, while less frequent, are responsible
for much of the topological changes.

% SEPARATION - FIG
\smallskip

%% Simulations saw that two well-separated particles may move
%% in opposite directions \cite{doliwa00}; however, we believe
%% these are the first observations of a nontrivial fraction of
%% nearest neighbors moving in opposite directions.

To investigate the nature of the correlated motion of
neighboring particles, we calculate the probability distribution
function, $P(\Delta R_{12}|R_{12})$, which measures the
change in separation, $\Delta R_{12}$, between two nearby
particles, initially separated by $R_{12}$, after a
time interval $\Delta t^*$.  These distributions are narrower
if $R_{12}$ corresponds to a peak of the pair correlation
function, $g(R_{12})$, than if $R_{12}$ corresponds
to a minimum, as shown by the solid and dashed lines in the inset
of Fig.~\ref{separation}(b), respectively.  This is shown more
dramatically in Fig.~\ref{separation}(b), where we compare
the widths of the distribution functions, $\sigma_{R_{12}}$,
directly with $g(R_{12})$, showing that they are anticorrelated.
This is reminiscent of the behavior of the collective diffusion
coefficient, which varies as the inverse of the static structure
factor, $D(q) \sim 1 / S(q)$~\cite{leshouches}.  The relaxation
of fluctuations at the peak of the structure factor, which
reflects the most favorable structure, is slowed relative
to other values of $q$.  Our results show a similar behavior
occurs in real space; particles whose separation corresponds to
a peak in $g(R_{12})$ are in more favorable relative positions
and tend to move together, so that their separation does not
change; by contrast, particles whose separation corresponds to
minima of $g(R_{12})$ are in less favorable relative positions,
and tend to move in antiparallel directions.

%% The symmetry of the distributions shown in the inset of
%% Fig.~\ref{separation}(b) suggests that converging and diverging
%% pairs of particles occur with equal probability.

%% commented out by Eric:
%% In fact, we examined several individual cases of particles
%% moving in anti-parallel directions and find that typically
%% these involve {\it four} particles: the particles start in a
%% diamond pattern around a small void, with two particles moving
%% in (to become nearest neighbors) while the other two move
%% outward to make room (becoming second-nearest neighbors). This
%% suggests a broad interpretation of the cooperativity of cage
%% rearrangements, ranging from ``string-like'' cooperative motion
%% to anti-parallel movements.

This work reveals a physical picture of cage-trapping and
rearrangement.  Cage rearrangements involve localized clusters
of particles with large displacements, in regions with higher
disorder and higher free volume.  Rearranging particles typically
move in parallel directions (``strings'').  However,
there are also ``mixing regions'' where
particles move in other directions [Fig.~\ref{separation}(a)],
accounting for a significant fraction of the
topological rearrangements.  Both the
collective nature of the relaxation, and the local origin of
the cage rearrangements, clearly play a key role in the behavior
of supercooled fluids of colloidal suspensions; it remains to be
seen whether they also are important effects in other glasses.

We thank J.~Conrad, J.~C.~Crocker, B.~Doliwa, U.~Gasser,
S.~C.~Glotzer, H.~Gould, and K.~Vollmayr-Lee for helpful
discussions.  We thank A.~Schofield for providing our colloidal
samples.  This work was supported by NSF (DMR-9971432) and NASA
(NAG3-2284).

%%%%%%%%%%%%%%%%%%%%%%%%%%%%%%%%%%%%%%%%%%%%%%%%%%%%%%%%%%%%
%%%               END OF THE PAPER                       %%%
%%%%%%%%%%%%%%%%%%%%%%%%%%%%%%%%%%%%%%%%%%%%%%%%%%%%%%%%%%%%

%% REFERENCES

% HELP \vfill\eject

% HELP \vfill\eject

% HELP \vfill\eject

% TABLE
\begin{table}
\caption{Estimates for the cage size, the cage rearrangement time
scale $\Delta t^*$, the cage lifetime $\Delta t^{**}$, and the
asymptotic diffusion coefficient $D_\infty$. }
\begin{tabular}{c | c c c c | c c | c}
% \hline
& \multicolumn{4}{c|}{cage size ($\mu$m)} &
\multicolumn{2}{c|}{time scales (hr)} \\
$\phi$ & $r_{\rm cage}$ & $r_{\rm free}$ &
$r_{\rm msd}$ & $\Delta r^{*}$ &
$\Delta t^*$ & $\Delta t^{**}$ & $D_\infty$ ($\mu$m$^2$/s) \\
\hline
0.46 & 0.75 & 0.55  & 0.63 & 1.12
& 0.083  & 0.52  & $15\cdot 10^{-5}$\\
0.52 & 0.35 & 0.35  & 0.23 & 0.40
& 0.17  & 2.1  & $0.80\cdot 10^{-5}$\\
0.53 & 0.45 & 0.31  & 0.27 & 0.49
& 0.67  & 9.4  & $0.30\cdot 10^{-5}$\\
0.56 & 0.25 & 0.21  & 0.17 & 0.29
& 0.28  & 3.3  & $0.26\cdot 10^{-5}$\\
% \hline
\end{tabular}
\label{cagetable}
\end{table}


\begin{thebibliography}{99}
\bibitem{footnote}
The distribution $P(\theta,\phi)$ remains qualitatively
unchanged even considering only pairs of particles with large
displacements, which are involved in cage rearrangements.
\end{thebibliography}
\end{document}